\documentclass[12pt,preprint]{aastex}

\slugcomment{To appear in Astron. J., September 2006}

\shortauthors{Fern\'andez et al.}
\shorttitle{Nucleus of Comet 162P}

\begin{document}

\title{Comet 162P/Siding Spring: A Surprisingly Large Nucleus}

\author{Y. R. Fern\'andez\altaffilmark{1,2}}
\affil{Dept. of Physics, Univ. of Central Florida,
4000 Central Florida Blvd., Orlando, FL 32816-2385}
\email{yan@physics.ucf.edu}

\author{H. Campins\altaffilmark{2}}
\affil{Dept. of Physics, Univ. of Central Florida,
4000 Central Florida Blvd., Orlando, FL 32816-2385; 
Lunar and Planetary Laboratory, Univ. of Arizona, 
1629 E. University Blvd.,
Tucson, AZ 85721}

\author{M. Kassis\altaffilmark{2}}
\affil{W. M. Keck Observatory, 65-1120 Mamalahoa Hwy, Kamuela, HI 96743}

\author{C. W. Hergenrother}
\affil{Lunar and Planetary Laboratory, Univ. of Arizona, 
1629 E. University Blvd.,
Tucson, AZ 85721}

\author{R. P. Binzel}
\affil{Massachusetts Institute of Technology, 77 Massachusetts Ave., 
Cambridge, MA 02139}

\author{J. Licandro}
\affil{Isaac Newton Group, P.O. Box 321, 38700, Santa Cruz de La Palma, 
Tenerife, Spain; Instituto de Astrof\'\i sica de Canarias, c/V\'\i a 
L\'actea s/n, 38205, La Laguna, Tenerife, Spain}

\author{J. L. Hora}
\affil{Harvard-Smithsonian Center for Astrophysics, 60 Garden St., 
MS 65, Cambridge, MA 02138-1516}

\author{J. D. Adams}
\affil{Dept. of Astronomy, Cornell University, Ithaca, NY 14853}

\altaffiltext{1}{Former address: Institute for Astronomy,
University of Hawaii, 2680 Woodlawn Dr., Honolulu, HI 96822.}
\altaffiltext{2}{Visiting Astronomer at the Infrared Telescope Facility, 
which is operated by the University of Hawaii under contract to the 
National Aeronautics and Space Administration.}

\begin{abstract}

We present an analysis of thermal emission from comet 162P/Siding
Spring (P/2004 TU$_{12}$) measured during its discovery apparition
in December 2004.  The comet showed no dust coma at this time, so
we have sampled emission from the comet's nucleus.  Observations
using the ``Mid-Infrared Spectrometer
and Imager" (``MIRSI") were performed at NASA's Infrared Telescope
Facility, where the peak of the comet's spectral energy distribution
was observed between 8 and 25 $\mu$m.  In combination with the three
near-infrared spectra presented by Campins {\it et al.} (2006, {\it
AJ}, this issue) that show the Wien-law tail of the thermal emission,
the data provide powerful constraints on surface properties of the
nucleus. We find that the nucleus's effective radius is $6.0\pm0.8$
km. This is one of the largest radii known among
Jupiter-family comets, which is unusual
considering the comet was discovered only recently.  Its geometric
albedo is $0.059\pm0.023$ in H-band, $0.037\pm0.014$ in R-band,
and $0.034\pm0.013$ in V-band.
We also find that the nucleus of 162P has little infrared beaming,
and this implies the nucleus has low thermal inertia. Including all
near-IR spectra yields a beaming parameter $\eta$ of $1.01\pm0.20$.
This result is in agreement with others showing that cometary nuclei
have low thermal inertia and little infrared beaming. If confirmed
for many nuclei, the interpretation of radiometry may not be as
problematic as feared.

\end{abstract}


%%%%%%%%%%%%%%%%%%%%%%%%%%
\section{Introduction}

A narrative of the discovery and morphological evolution of comet
162P is given by \citet{cam06} in a companion paper (hereafter Paper
I).  The thin tail and lack of coma strongly suggested that
observations of this comet would reveal flux from the nucleus.
Since opportunities to directly study a cometary nucleus are
infrequent, our group set up a multiwavelength observing campaign
to take advantage of the comet's unusual behavior and probe the
reflectance and thermal properties of the nucleus.  Paper I addressed
the near-infrared spectroscopic properties; in this paper we describe
the thermal emission and the albedo, which include observations
from the Wien-law tail through the black-body peak in wavelength.

%%%%%%%%%%%%%%%%%%%%%%%%%%
\section{Observations and Reduction}

The mid-infrared observations were obtained at NASA's Infrared
Telescope Facility  between UT 0600 and 0700 on December 27, 2004,
using the ``Mid-Infrared Spectrometer
and Imager" -- ``MIRSI" -- instrument \citep{deu02}. Conditions
were clear and photometric, but only moderately dry.
The comet was 1.348 AU from
the Sun, 0.777 AU from Earth, and at a phase angle of 46.1 degrees.
Chopping and nodding were employed in such a way as to leave four
images of the comet on the detector, effectively boosting our
signal-to-noise by a factor of 2. The comet appeared as a point-source
and there is no chance that we were chopping onto coma.
The comet was located just 12.7$^\circ$ away from 
the standard star $\beta$ And in the sky, which was used as an absolute
flux calibrator. During its observations, 
the airmass of $\beta$ And was 1.05 to 1.14. During
the observations of 162P,  the airmass was 1.14 to 1.19 and
then later 1.31 to 1.36. 
Both comet and star were observed at four wavelengths,
8.7, 9.8, 11.7, and 24.5 $\mu$m.  The filters through which
we observed the comet and the star at significantly different
airmasses ($\sim$0.2) were 8.7 and 11.7 $\mu$m. However, this airmass
difference appeared to be not important:
the 11.7 $\mu$m flux density was measured at the two airmass values
and not found to differ significantly. Since 8.7 $\mu$m
is a relatively clean part of the N-band, we applied no airmass
correction. Given the uncertainty of the
photometry (below), this assumption is valid.

MIRSI's filters at the four wavelengths are about 10\%
wide.  We assumed that $\beta$ And had flux densities of 351.0,
279.3, 198.2, and 46.0 Jy at the four wavelengths, based on the
results of \citet{coh96} using the functions of \citet{eng92}.
Aperture photometry of the comet yielded instrumental magnitudes,
which we corrected for aperture size. We calculated
absolute flux densities for the comet of $2.0\pm0.4$ Jy, $2.3\pm0.4$
Jy, $2.6\pm0.4$ Jy, and $2.5\pm0.6$ Jy at the four wavelengths.

In addition to our mid-IR data we make use of the near-IR spectra
described in Paper I. The spectra were obtained at
December 3.25, 10.25, and 11.88 of 2004 (UT).
Also we use visible-wavelength photometry
published by \cite{herg06}. Those observations were obtained
on November 19.16 and December 9.19 of 2004,
and April 6.15 of 2005 (all UT).

%%%%%%%%%%%%%%%%%%%%%%%%%%
\section{Analysis}

The near-infrared spectra were not flux-calibrated on an absolute
scale.  Hence we used a two-pronged approach in analyzing the data.
Step 1 was to attack just the {\sl absolute} photometry in the
mid-IR and visible wavelengths. Step 2 was to analyze just the {\sl
relative} fluxes in the near-IR spectra.  We describe these below.

\subsection{Absolute Photometry}

The basic radiometric method to obtain an effective radius, $R$, and
geometric albedo, $p$, is to solve two equations with these two
unknowns, first done about 35 years ago \citep{all70,mat72,mor73} and
described in detail by \citet{ls89}:
\begin{mathletters}
\begin{eqnarray}
F_{vis}(\lambda_{vis}) & = & 
                {{F_{\odot}(\lambda_{vis})}\over{(r/1{\rm AU})^2}}\ R^2 p
                                 {{\Phi_{vis}(\alpha)}\over{\Delta^2}}, \\
F_{mir}(\lambda_{mir}) & = & 
        \epsilon\!\int\!B_\nu(T(pq,\eta,\epsilon,\theta,\phi),\lambda_{mir}) 
                d\phi d\cos\theta\ R^2\ {{\Phi_{mir}(\alpha)}\over{\pi\Delta^2}
},
\end{eqnarray}
\end{mathletters}
where $F$ is the measured flux density of the object at wavelength
$\lambda$ in the visible (``vis'') or mid-infrared (``mir'');
$F_{\odot}$ is the flux density of the Sun at Earth as a function
of wavelength; $r$ and $\Delta$ are the object's heliocentric and
geocentric distances, respectively; $\Phi$ is the phase darkening
in each regime as a function of phase angle $\alpha$; $B_\nu$ is
the Planck function; $\epsilon$ is the infrared emissivity; $\eta$
is a factor to account for infrared beaming; and $T$ is the
temperature. The temperature itself is a function of $p$, $\epsilon$,
$\eta$, surface planetographic coordinates $\theta$ and $\phi$, and
the (dimensionless) phase integral $q$ which links the geometric
and Bond albedos.  For lack of detailed shape information, the
modeled body is assumed to be spherical, so all radii given here
are ``effective'' radii.

The surface map of temperature is calculated using a model of the
thermal behavior. A simple thermal model covering an extreme of
thermal behavior is often employed \citep{ls89}.  This model, the
``standard thermal model'' (STM), is widely used so results are
easy to compare. It applies if the rotation is so slow or the thermal
inertia so low that every point on the  surface is in instantaneous
equilibrium with the impinging solar radiation.  In this case the
temperature is a maximum at the subsolar point and decreases as
$\root 4 \of {\cos\vartheta}$, where $\vartheta$ is the local solar
zenith angle. We will show below that the modeling results are
consistent with this assumption.

The other parameters to the model are $\epsilon$, $\Phi_{mir}$,
$\Phi_{vis}$, $q$, and $\eta$. We have sidestepped the need to
explicitly assume values of $\Phi_{vis}$ and $q$ by using the
nucleus's absolute magnitude $H$ (see below).  Emissivity for rocks
is close to unity \citep{mor73} and we assume $\epsilon=0.9$ here.
The choices of $\Phi_{mir}$ and $\eta$ are often the most important
determinants in deriving an accurate $R$ and $p$.  We have used the
NEA Thermal Model (NEATM) devised by \citet{har98} to determine
what to use for these quantities. This model's characterization of
$\Phi_{mir}$ and $\eta$ is the primary distinguishing difference
from the pure STM.  For $\Phi_{mir}$, \citet{har98} argued that a
more sophisticated phase law is needed rather than the usual linear
phase coefficient. His approach is to calculate a phase effect based
simply on the surface-integral of the thermal flux over the
Earth-facing hemisphere.

As for $\eta$, the standard value of $0.756$ \citep{leb86} was
originally derived for Ceres. However recent work
\citep{har98,hdg98,hd99,del03} indicates that small bodies can have
a variety of values for $\eta$.  Thus we have made the beaming
parameter a variable to be fit, and our modeling routines return
values for three physical parameters, $R$, $p$, and $\eta$.

We have four mid-IR photometry points, but we need at least one
reflected-sunlight measurement. Ideally this measurement would have
taken place simultaneously, but unfortunately no such data were
taken.  Instead we make use of R-band photometry published by
\cite{herg06}.  With their measurements taken over a range of phase
angles, we derive an R-band absolute magnitude $H_R$ of $13.74\pm0.25$
and a phase slope parameter $G$ of $0.15\pm0.10$ \citep{bow89}.
With this photometric constraint, we now effectively have five total
measurements to fit three parameters, which leaves us with two
degrees of freedom.

The results of the modeling are given in Figs. 1 and 2.  Figure 1
displays contour plots for three representative values of $\eta$:
0.75, 1.0, and 1.25. The contours trace out the 1, 1.5, 2, 2.5, and
3-$\sigma$ levels for $\chi^2_\nu$.  We found the full 1-$\sigma$
ranges of the three parameters to be as follows:  3.4 km $< R <$
6.8 km, 0.022 $< p_R <$ 0.102, and 0.30 $< \eta < 1.29$. These are
fairly wide and poorly constrained intervals, but we note that the
values are highly correlated. This is demonstrated in Fig. 2, where
the best-fitting values of $R$ and $p_R$ are plotted for a given
value of $\eta$.  Also shown is the correlation between $R$ and
$p_R$.  This figure means that if we can  use the near-infrared
spectra to place another constraint on $p$ or $\eta$, we will be
able to very tightly constrain the range of acceptable values.

\subsection{Relative Spectral Behavior}

As described by \cite{riv05} and \cite{abe03}, JHK-band spectroscopy
of a sufficiently-hot object in the inner Solar System will show
the Wien-law tail of the object's thermal emission. The curvature
of the spectrum in the transition between the reflected component
and the thermal component depends on the object's geometric albedo
$p$ and the beaming parameter $\eta$.  (Since both components depend
on $R^2$, radius does not matter.) The more (less) reflective the
object, the longer (shorter) the wavelength at which the transition
occurs.  Similarly, the higher (lower) the value of the infrared
beaming parameter, the cooler (hotter) the subsolar point is, and
the longer (shorter) the wavelength at which the transition occurs.
Note in particular that the nucleus's Wien-side 
of its spectral energy distribution (SED) is very sensitive
to the emission from the subsolar point since that is the hottest
location on the object.  Also, note that generally the subsolar
point is hotter than any dust that may reside in the coma, since
dust grains are isothermal and often (in Jupiter-family comets) do
not have significant superheat \citep{lis02}.  Therefore the comet's
Wien-side SED is not as susceptible to coma contamination.

The three spectra presented in Paper I are reproduced in our Fig.
3.  Each panel shows one spectrum and modeling results (described
below).  The spectra have had the solar spectrum divided out. It
is clear that the color of the nucleus is not flat, i.e. not solar,
so the nucleus's geometric albedo depends on wavelength.  In our
case, the most robust albedo to be obtained from our modeling here
is the H-band albedo, $p_H$.  This is because the three spectra
show a kink near 1.3 $\mu$m and have a constant sloping trend with
wavelength only longward of that.  Trying to model the kink would
not help us constrain the thermal properties of the nucleus; it is
likely a real feature and not due to systematic observational
problems, but at this point we do not need to understand its origins
in order to continue our analysis. We defer further study of this
spectral feature to future work.  From Paper I (their Fig. 2) we
can derive an estimate of the relative reflectance between R-band
and H-band: the albedo at H is about 60\% higher than at R.  Thus
$p_H$ must be divided by 1.6 to yield $p_R$.

Since the spectra are not flat and reflectivity increases with
wavelength, we must incorporate a ``reddening" slope into the
modeling. To determine the appropriate reddening, for each spectrum
we fit a linear slope through the data between wavelengths of 1.3
and 1.9 $\mu$m. (This avoids the kink and the thermal emission.)
We find that the slope is about $0.30\pm0.02$ per micron in the
December 3rd spectrum, $0.28\pm0.02$ in the December 10th spectrum,
and $0.39\pm0.03$ in the December 11th spectrum. This assumes a
reflectance of unity at 1.2 microns.  We incorporated these slopes
and their uncertainties into the modeling; i.e., we tried various
values of the slopes to be sure that the error budget of our final
results included an accounting for the uncertainty in the reddening.

We have a two-parameter model with which to analyze the spectra,
and the two parameters are $p_H$ and $\eta$. The model is similar
to that described in the previous subsection. Equation 1b is used
to derive the thermal emission. Equation 1a is used for the reflected
sunlight but modified to account for the reddening mentioned above;
the solar spectrum is simply reddened by the appropriate amount as
a function of wavelength (with no reddening at 1.6 $\mu$m).  To
account for phase angle, we set $G=0.15$, as derived above, and we
incorporated this as $\Phi_{vis}$.  Then an overall solar spectrum
is divided out to match the presentation in Fig. 3. We modeled each
of the three spectra individually, using only wavelengths between
2.0 and 2.4 $\mu$m.  (Shortward of that, there is no extra information
for us; longward of that, the data could be unreliable.) The
uncertainty in the reflectance of the December 3rd spectrum was
about 0.5\% in each spectral bin; of the December 10th spectrum,
about 1\% in each spectral bin; and of the December 11th spectrum,
about 1.5\% in each spectral bin.  We used these errors in calculating
the fit statistic $\chi^2_\nu$.

Some example results are shown in Fig. 3, where we have chosen the
best fitting value of $p_H$ for three values of $\eta$: 0.8, 1.0,
and 1.2. As is apparent in the figure, each of these example models
adequately fits its respective spectrum.  Note that there is good
agreement between the model results of the December 3rd and 10th
spectra. The results for the December 11th spectrum are somewhat
disparate.  This could be physically related to the higher reddening
slope, but it is likely that the softer curve of the spectrum into
the thermal tail is an important effect as well.  In any case, as
we discuss below, there is no {\sl a priori} reason to disregard
the December 11th results, and the differences do not adversely
effect our final results too much.

\subsection{Synthesis}

The two analysis methods give ranges of $p$ and $\eta$ that intersect
in parameter space. By combining the best fits from both methods
we can narrow down the allowable range of $p_H$, $p_R$, and $\eta$.
This is shown in the top panel of Fig. 4, where we plot the $1\sigma$
contours of $p_H$ and $\eta$ phase space derived from the near-IR
spectral analysis.  Each spectrum has its own contour. Note
how there is overlap between the results for the three spectra
(although there is no single location itself overlapped by all three).
Overplotted
is the $1\sigma$ contour from Fig. 2, but with the albedo scaled
up by 60\% to account for the difference in R- and H-band albedos.
The true albedo and beaming parameter should fall within the overlap
of the mid-IR and near-IR results.

This overlap is shown in the bottom panel of Fig. 4. The overlap
between the mid-IR contour and each individual near-IR spectrum
is shown with a shaded region. The combination of all three shaded
regions gives us our answer for the beaming parameter and
albedo. We find that $\eta=1.01\pm0.20$ and 
$p_H=0.059\pm0.023$. This corresponds to 
$p_R=0.037\pm0.014$.
For reference, from Fig. 2 in Paper I we find that $p_R$ is 9\%
higher than $p_V$, therefore $p_V=0.034\pm0.013$.
Using our Fig. 2, and the allowed range of $\eta$, the effective radius
of 162P's nucleus is $R=6.0\pm0.8$ km. 

%%%%%%%%%%%%%%%%%%%%%%%%%%
\section{Discussion}

\subsection{Absolute Magnitude and Phase Slope Parameter}

As mentioned, we have derived a value for the R-band
absolute magnitude $H_R$  based on photometry
that covers a range of phase angles. However the rotational context
is not known, and for a proper analysis of the phase law, one needs
to account for the variability of the magnitude due to the nucleus's
rotation. Our lack of this information introduces an uncertainty
into our derivation of $H_R$ and $G$. There is a possibility that
if the nucleus's light curve amplitude is large then we are being
fooled, and these two quantities may have quite different values.

In particular this is a concern because a slope parameter of $G=0.15$
is somewhat higher than has been found for several other cometary
nuclei and outer Solar System objects \citep{fer00,sr02, sj02,bau03,js04},
although it is comparable to that found by \cite{bur04} for comet
19P/Borrelly.  The problem is compounded by the fact that the mid-IR
and near-IR observations both took place at moderate phase angle
($46^\circ$ and $49^\circ$).  If 162P's slope parameter is indeed
much smaller than $0.15$, this would mean that $H_R$ is brighter
than we have calculated.  This, in turn, would mean that the nucleus's
true albedo is higher and effective radius
is lower than we have derived here.  For example, if
the nucleus's $G=-0.05$, then the infrared observations would imply
that $p_R\approx0.06$ and $p_H\approx0.09$.  Such an albedo would
be somewhat higher than the other albedos known for Jupiter-family
comets \citep{lam05}, and be closer to those of Centaurs and
transneptunian objects \citep{stan05}. Furthermore the 
effective radius would be $R\approx4.7$ km, admittedly still
a fairly large nucleus.

\subsection{December 11th Spectrum}

The fact that the December 11th spectrum is somewhat different
from those of the other dates suggests that we should explore
why this may be. First we can
check the rotational context of the observations.
The rotation period of the comet is about 32.78 hours
(G. Masi, private communication), so we can calculate the relative
rotational phase between the four epochs. If the December 3rd
spectrum occurred at rotational phase $\phi=0$, then at the other
epochs, $\phi=0.13$ for the December 10th spectrum, $\phi=0.32$ for
the December 11th spectrum, and $\phi=0.59$ for the December 27th
photometry.  Therefore there is no clear conclusion we can draw
about what effect the rotation may have had on our interpretation
of the modeling results. For example
if the December 11th spectrum had been at
a much different rotational phase compared to the other three
datasets, then we would have reason to disregard it in the final
analysis, but this is not the case. (We admit that if the rotation
period is wildly off from 32.78 hours then we will have to readdress
this argument.) In any case the lack of rotational context for the
visible-wavelength photometry introduces extra uncertainty in the
problem.  

Another explanation may be secular changes in the comet's
activity. For example, perhaps there was
an outburst of activity sometime between December 10.25 and December
11.88.  This would have formed a dust coma that would affect the
spectral measurements. If the dust grains had sizes that were
approximately 1 to 10 $\mu$m in scale, then they would have contributed
extra reflected sunlight but little thermal emission in K-band.
This is because the grains would have been isothermal and would not
have reached the hotter temperatures that are achieved at and near
the nucleus's subsolar point. Modeling such a spectrum on the
assumption that we were only seeing light from the nucleus would
yield higher albedos than normal -- which is exactly what is seen
in the December 11th spectrum.  If this scenario is correct, then
we should not include this spectrum in the derivation of the
properties of the nucleus.

Finally, we note that the signal-to-noise ratio ($S/N$) of the spectrum on
December 11th is somewhat 
lower than that from December 3rd and 10th. The $S/N$
of this spectrum is still good and cannot be rejected for that reason,
but if we were to take just the two higher $S/N$ spectra
we would find that the acceptable range of $\eta$
is $0.98\pm0.17$ (as can be found from
Fig. 4). This is not too different from our
solution in \S 3.3.  Furthermore there would be no difference
in our range of acceptable albedos $p_H$ (as can
also be seen from Fig. 4). 
The acceptable range of radii $R$ would shrink by only
a few tenths of a kilometer (as can be seen from Fig. 2). 

In short,  there is no conclusive evidence letting
us explain why the December 11th spectrum is different,
and also no strong reason why it should be removed from
consideration.  

\subsection{Thermal Properties}

Only recently have there been sufficient thermal measurements of
cometary nuclei that can constrain the thermal inertia.  In particular,
the thermal emission from the nucleus of 9P/Tempel 1 strongly
indicates that the beaming parameter $\eta$ is near unity, and that
the thermal inertia is consistent with zero. This comes from both
remote observations by \cite{lis05} and from spacecraft observations
reported by \cite{ahe05} from the {\sl Deep Impact} mission.
\cite{sod04} analyze spatially-resolved {\sl Deep Space 1} near-IR
spectra of comet 19P/Borrelly's nucleus and find that the nucleus's
temperature is consistent with a similar thermal behavior.  Lastly,
observations of comet 2P/Encke by Fern\'andez et al. (2004; and in
preparation) indicate that that nucleus has a near unity beaming
parameter as well. This is based on mid-IR, near-IR, and visible
observations, just as in our present analysis of 162P.

If cometary nuclei are confirmed to commonly have low thermal
inertias and near-unity beaming parameters, this will greatly
facilitate the interpretation of ground-based mid-IR radiometry.
For faint comets often one can detect the comet at only one mid-IR
wavelength. This is insufficient to constrain $\eta$ and so in such
a case one needs to assume a value for it. If we know {\sl a priori}
that there is a strong likelihood of $\eta=1$, then this problem
is removed.  Furthermore one would not necessarily need to obtain
time-consuming photometry at many extra wavelengths. A survey of
the thermal emission from cometary nuclei would be logistically
easier and would yield significant physical information with
relatively few model-dependent problems.

We note that our solution for $\eta$ is fairly low given the phase
angle of the observations.  \cite{del03} found an approximately
linear relationship between the beaming parameters and phase angles
of several near-Earth asteroids (NEAs). Their best-fit line would
suggest that $\eta$ at $46^\circ$ should be near 1.27. However there
is much scatter in their trend and our solution here for 162P is
not significantly off. In any case, cometary nuclei likely have
different thermal properties anyway.  The fact that we do not see
an elevated $\eta$ at such phase angles, unlike what has been seen
for many NEAs, is suggestive of this.

\subsection{Radius and Albedo Context}

A large database of physical information on cometary nuclei was
compiled by \cite{lam05}.  By comparing with this work, our radius
for 162P makes it  one of the largest Jupiter-family comets (JFCs).
Only 28P/Neujmin 1 has an accepted radius that is larger, and
10P/Tempel 2 and 143P/Kowal-Mrkos have comparable radii. This is
significant since it was only discovered in October 2004, despite
its large size, so perhaps the sample of known JFCs is not as
complete as previously thought. Studies of the JFC size distribution
by \cite{jf99}, \cite{wl03}, \cite{mhm04}, and \cite{lam05} suggested
that we knew about all the nuclei larger than a few kilometers, or
at least those with perihelia within roughly 2 AU of the Sun.  162P's
perihelion is at 1.2 AU, and its current
orbit-intersection
distance with Earth is 0.23 AU, so it is a fairly close yet large comet
that avoided discovery.  While this is partly due to poor apparitions,
it also suggests that the comet's weak level of cometary activity
-- as shown in late 2004 and early 2005 -- kept it below the threshold
brightness for discovery. The question of our completeness of the
JFC sample perhaps needs to be revisited.

The albedo of the nucleus appears to be typical. While the error
bar on the albedo is approximately 40\%, it overlaps with a significant
fraction of the known albedos among JFCs \citep{lam05}. However,
it is important to point out that there are only nine other albedos
known for JFCs, and that we do not yet have a statistically robust
sample with which to make a mathematical comparison.

%%%%%%%%%%%%%%%%%%%%%%%%%%%%%%%%%%%%
\section{Summary}

Mid-infrared observations in December 2004 have let us sample the
thermal emission from the nucleus of comet 162P/Siding Spring. In
combination with visible-wavelength photometry reported by \cite{herg06}
and three near-infrared spectra reported by \cite{cam06} (Paper I),
we have calculated the nucleus's effective radius $R$, geometric
albedo $p$, and beaming parameter $\eta$.

We used a two-pronged analysis where we constrained all three
quantities with the mid-IR and visible data, and then independently
constrained $p$ and $\eta$ with the near-IR spectra.  The overlap
of the ranges of acceptable ($1\sigma$) values for these quantities
lets us derive a beaming parameter of $\eta=1.01\pm0.20$, and a
geometric albedo of $p_H=0.059\pm0.023$ and $p_R=0.037\pm0.014$ in
H- and R-band respectively.  We also find that the effective radius
at the time of the December 27th observations was $6.0\pm0.8$ km.

There are two significant conclusions. First, we find that the
nucleus is one of the largest known among Jupiter-family comets.
This suggests that we must be cautious in understanding the
completeness level to which we have discovered all JFCs. Second,
with a beaming parameter near unity, this continues a trend of
finding little infrared beaming among cometary  nuclei.  If this
property is widespread, radiometric observations of cometary nuclei
will be much easier to interpret and less prone to model dependencies.

%%%%%%%%%%%%%%%%%%%%%%%%%%
\acknowledgments{
We are indebted to the support team at IRTF for making these
observations possible. Our work benefited from the very thorough
analysis by an anonymous referee.  We acknowledge the JPL SSD group
for their very useful ``Horizons'' ephemeris service. This work was
supported in part by a SIRTF Fellowship to YRF. HC acknowledges
support from grants from NASA's Planetary Astronomy program and
from the National Science Foundation.}

%%%%%%%%%%%%%%%%%%%%%%%%%%

\clearpage

%%%%%%%%%%%%%%%%%%%%%%%%%%
% Figures

\clearpage

\begin{figure}
\epsscale{0.45}
\plotone{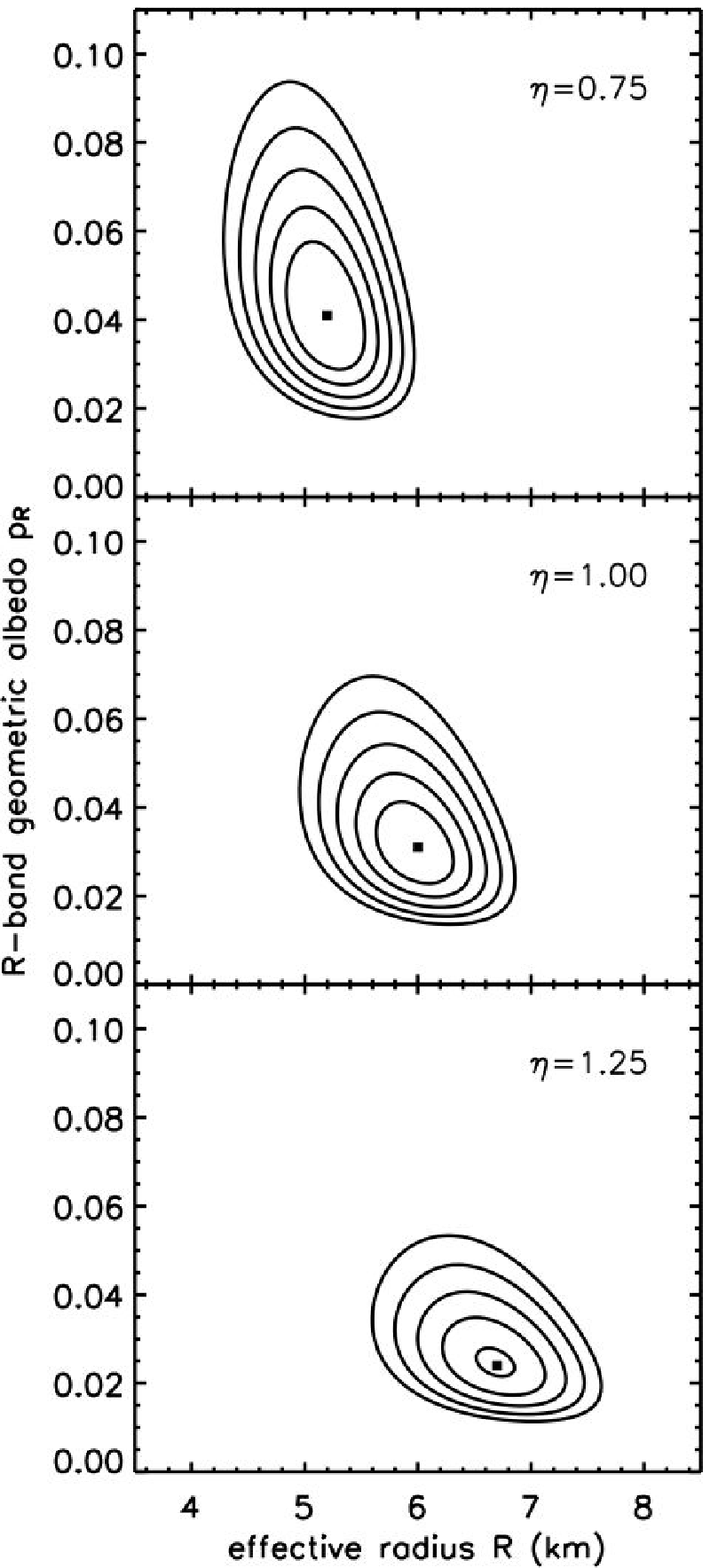}
\caption{%
Contour plots of the fit statistic $\chi^2_\nu$ after
modeling the mid-IR photometry and visible absolute magnitude. We
have represented the three-dimensional nature of $\chi^2_\nu$ by
showing three representative values of the beaming parameter $\eta$:
0.75, 1.0, and 1.25. In each plot, the five contours represent the
1, 1.5, 2, 2.5, and 3-$\sigma$ contours. The filled square shows
the lowest $\chi^2_\nu$ for that particular value of $\eta$.  A
wide of range of $R$, $p_R$ and $\eta$ are viable solutions.}
\end{figure}
\clearpage

\begin{figure}
\epsscale{0.45}
\plotone{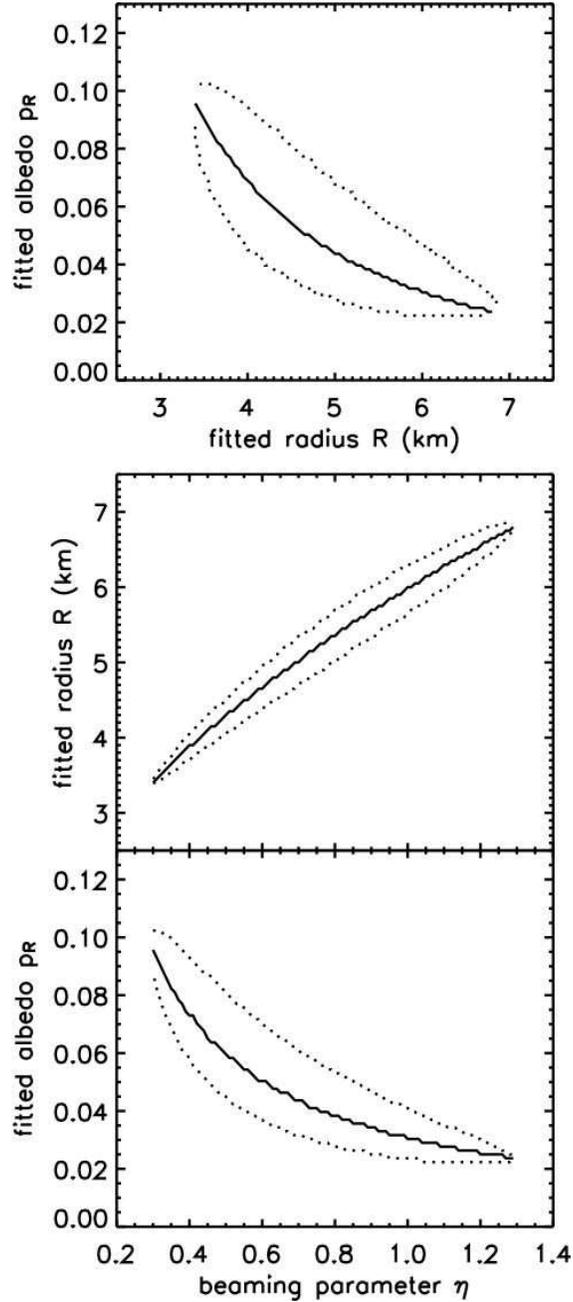}
\caption{Plots showing the correlation between the three parameters
that were fit, $R$, $p_R$, and $\eta$. Dotted lines represent the
1-$\sigma$ ranges. The jagged
nature of the curves is due to the quantization of the sampling of
parameter space.  While Fig. 1 shows that each parameter individually
has a wide range of possible values, the plots here show that the
values are strongly correlated.  Another constraint on any one of
these parameters will yield a dramatic decrease in the error bar
associated with all three.}
\end{figure}
\clearpage

\begin{figure}
\epsscale{1.0}
\plotone{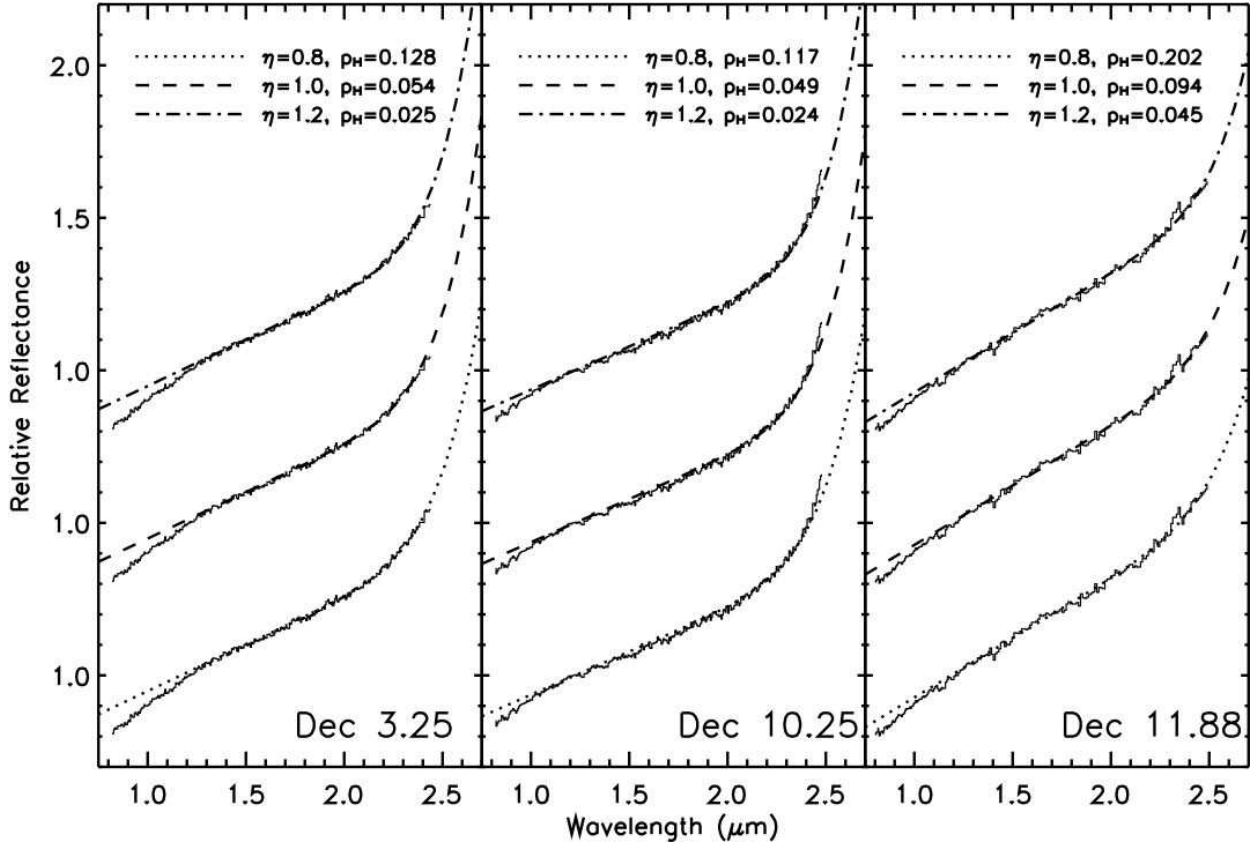}
\caption{Model results of the three spectra presented in Paper I.
Each panel represents one spectrum, with the date of that spectrum
given at bottom right.  Three example models are shown for each
spectrum, and the spectrum has been offset from itself for clarity.
The model parameters, $\eta$ and $p_H$ are written at top. 
To calculate the fit statistic for a model, we used an error bar
of 0.5\%, 1\%, and 1.5\% on the reflectance of each spectral bin
in the December 3rd, 10th, and 11th spectra, respectively.  As with
the mid-IR analysis, a wide range of values satisfy the spectra.
Note however that generally higher albedos are needed for the
December 11th spectrum.}
\end{figure}
\clearpage

\begin{figure}
\epsscale{0.45}
\plotone{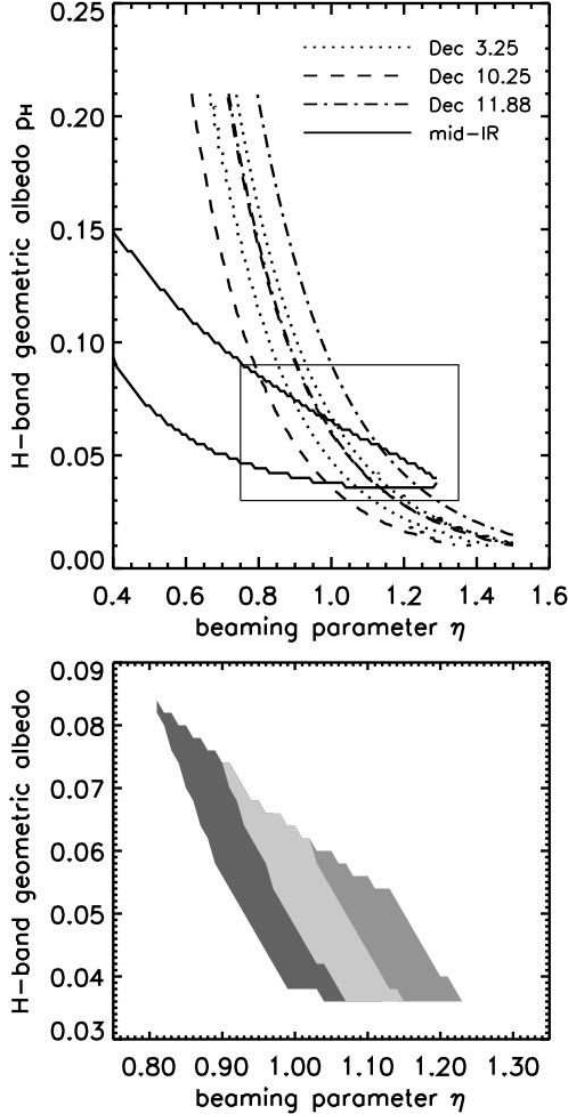}
\caption{Top: Contour plot of the $1\sigma$ boundaries for $p_H$
and $\eta$ as derived from fitting the three near-IR spectra.
The model results for each spectrum are shown. The parameter
spaces of the three spectra overlap.  Overplotted is the $1\sigma$
region from Fig. 2 showing the mid-IR constraint on $p_H$ and $\eta$.
We have scaled up $p_R$ by 60\% in order to compare it to the $p_H$
here. The rectangle shows the area encompassed by the bottom panel.
Bottom: Shaded regions showing various overlaps, showing how
we derive the allowable ranges to $p$ and $\eta$. The lightest,
darkest, and middle grey indicate overlap between the mid-IR results
and the spectrum results from December 3rd, 10th, and 11th, respectively
(though the December 3rd region is on top, covering some of the
overlaps of the other two dates). Our estimate of the nucleus's
$\eta$ and $p_H$ comes from the extent of this entire shaded region.}
\end{figure}
\clearpage

\end{document}